\documentclass[sigconf]{acmart}
\acmConference[MSR 2024]{21st International Conference on Mining Software Repositories}{April 2024}{Lisbon, Portugal}

\usepackage{tikz}
\usepackage{amsmath}
\usepackage{filecontents}
\usepackage[font=small,skip=0pt]{caption}
\usepackage{multirow}
\usepackage{csquotes}
\usepackage{placeins}
\usepackage{xspace}
\usepackage[normalem]{ulem}
\usepackage{soul}
\usepackage{url}
\usepackage{enumitem}
\usepackage{hyperref}
\usepackage{bbding}
\usepackage{pifont}
\usepackage{wasysym}
\usepackage[utf8x]{inputenc}
\usepackage{subcaption}
\usepackage[most]{tcolorbox}

\newcommand{\ie}{\emph{i.e.,}\xspace}
\newcommand{\eg}{\emph{e.g.,}\xspace}
\newcommand{\etal}{\emph{et al.}\xspace}
\newcommand{\cut}[1]{}

\usepackage{xstring}
\newcommand{\PP}[1]{
\vspace{2px}
\noindent{\bf \IfEndWith{#1}{.}{#1}{#1.}}
}

\newcommand{\squishlist}{
  \begin{list}{$\bullet$}
    { \setlength{\itemsep}{0pt}      \setlength{\parsep}{3pt}
      \setlength{\topsep}{3pt}       \setlength{\partopsep}{0pt}
      \setlength{\leftmargin}{1.0em} \setlength{\labelwidth}{1em}
      \setlength{\labelsep}{0.5em} } }
\newcommand{\squishend}{
    \end{list}  }

\def\Snospace~{\S{}}

\renewcommand\footnotetextcopyrightpermission[1]{}

\begin{document}

\title{An Investigation of Patch Porting Practices of the \\Linux Kernel Ecosystem}

\author{Xingyu Li}
\affiliation{%
  \institution{UC Riverside}
}
\email{xli399@ucr.edu}

\author{Zheng Zhang}
\affiliation{%
  \institution{UC Riverside}
}
\email{zzhan173@ucr.edu}

\author{Zhiyun Qian}
\affiliation{%
  \institution{UC Riverside}
}
\email{zhiyunq@cs.ucr.edu}

\author{Trent Jaeger}
\affiliation{%
  \institution{UC Riverside}
}
\email{trent.jaeger@ucr.edu}

\author{Chengyu Song}
\affiliation{%
  \institution{UC Riverside}
}
\email{csong@cs.ucr.edu}

\begin{abstract}

Open-source software is increasingly reused, complicating the process of patching to repair bugs. In the case of Linux, a distinct ecosystem has formed, with Linux mainline serving as the upstream, stable or long-term-support (LTS) systems forked from mainline, and Linux distributions, such as Ubuntu and Android, as downstreams forked from stable or LTS systems for end-user use. Ideally, when a patch is committed in the Linux upstream, it should not introduce new bugs and be ported to all the applicable downstream branches in a timely fashion. However, several concerns have been expressed in prior work about the responsiveness of patch porting in this Linux ecosystem. In this paper, we mine the software repositories to investigate a range of Linux distributions in combination with Linux stable and LTS, and find diverse patch porting strategies and competence levels that help explain the phenomenon. Furthermore, we show concretely using three metrics, i.e., patch delay, patch rate, and bug inheritance ratio, that different porting strategies have different tradeoffs. We find that hinting tags(e.g., Cc stable tags and fixes tags) are significantly important to the prompt patch porting, but it is noteworthy that a substantial portion of patches remain devoid of these indicative tags. Finally, we offer recommendations based on our analysis of the general patch flow, e.g., interactions among various stakeholders in the ecosystem and automatic generation of hinting tags, as well as tailored suggestions for specific porting strategies.

\end{abstract}

\maketitle

\section{Introduction}

Open-source software has greatly lowered the cost of software development.
It is a common practice to reuse, customize, and even rebrand open-source projects.
This also means that multiple versions or forks of the same upstream project can co-exist.
In this study, we focus on the Linux kernel ecosystem. It is widely reused in numerous desktop and server distributions like Ubuntu, Fedora, and Red Hat, as well as in billions of Android, Amazon AWS, and Internet of Things (IoT) devices.
These Linux forks are often based on specific Linux stable or long-term support (LTS) branches of the kernel source, which in turn were originally forked from the Linux mainline (also sometimes called Linux upstream). 
The Linux mainline is where all the development and bug fixes occur --- including feature commits and patch commits. 
In contrast, stable or LTS branches only port patch commits from mainline, as their goal is to maintain stability.

For large and complex open-source projects like Linux that are being actively developed, it is inevitable that there will be a continuous stream of bugs being discovered (and introduced), including security vulnerabilities.
Unfortunately, the co-existence of many forked downstream branches makes it challenging to keep track and fix issues in a timely manner.
In Linux specifically, all bugs are supposed to be fixed in the mainline first (a.k.a., upstream first)~\cite{patch-apply-process}.  
Once a patch is merged into the mainline, downstream kernel maintainers are responsible for porting patches from upstream.
But, as we will show, it is often not straightforward to figure out which of the upstream patches should be ported, especially considering that downstream branches may be forked from older versions that have diverged from the upstream, requiring more complex backporting. 
This problem is further exacerbated because the downstream Linux distributions are maintained by independent organizations with differing priorities.

Ideally, all of downstream branches, including Linux distributions and Linux stable/LTS branches should port all necessary and applicable patches from Linux mainline, and do not introduce any new bugs at the same time. 
However, we know anecdotally that the patch porting process is far from ideal.
Several studies have been conducted to measure the vulnerability lifetime and patch timeliness~\cite{li2017large,jones2020deploying,farhang2019hey,shahzad2012large, frei2006large},
though few studies~\cite{zhang2021investigation} focus on an ecosystem with upstream and downstream kernels.
We believe a study of existing patch porting practices or strategies employed by the individual parties in the ecosystem is beneficial in several ways. For example, we hope to understand the pros and cons of different patch porting strategies, evaluate their implementation, and identify opportunities for enhancing the patch process in the Linux ecosystem (some of which may generalize to other open-source software at large).

To conduct this research, we utilized various publicly available data sources to produce measurements, which we then form hypotheses. To validate the hypotheses, we also engage in communication with 23 Linux distributions and LTS maintainers, who are among the top contributors in public Linux git repositories. In addition to measurement results, we also relied on (1) publicly available documents and (2) reverse-engineering labels in commit messages, which were then verified through discussions with the maintainers.

For measurements, we focus on three main metrics: (1) patch delay: how long it takes a patch to port from upstream to downstream; (2) patch rate: the frequency of patch porting, i.e., number of ported patches per unit time; (3) bug inheritance ratio: the fraction of mainline bugs introduced in a downstream branch (in a given time period). In addition, to understand patch practices and implications better, we also categorize patches based on various information, e.g., security patches, labels provided by patch authors.

At a high level, we are interested in these high-level questions:
\begin{enumerate}[label=\textbf{RQ{\arabic*}}]
    \item What are the patch porting strategies employed by downstream kernel maintainers? 
    \item What is the performance of these strategies in terms of the three metrics we defined? How far away are they from ideal?
    \item What improvements can be made to enhance the patch porting practices?
\end{enumerate}

Overall, by studying 21 branches across 8 popular Linux distributions (including Android), we find drastically different patch porting strategies are employed by the various downstream kernel branches,
and none of the strategies can achieve best results in all three metrics simultaneously.
Specifically, if they choose to favor porting as many patches as possible and as quickly as possible, then they unavoidably
introduce many more bugs at the same time. Conversely, if they choose to favor stability (i.e., fewer bugs introduced), then 
they tend to miss patches or port them with significant delays. 
Interestingly, even for the downstream branches that do favor stability, their patch porting performance can still differ drastically, indicating that there is significant room for improvement.

We summarize our contributions as follows:

\squishlist
\item \cut{We explore the Linux kernel ecosystem to measure its patch porting delays and find that there are excessive delays between Linux mainline and Linux LTS/distributions.}
We investigate patch porting strategies in the Linux ecosystem both qualitatively and quantitatively. Specifically, we define three metrics that comprehensively and objectively capture the performance of patch porting strategies.

\item 
In our study, we do not only report objective metrics but also look for factors that contribute to the performance captured by the metrics. This helps us understand the operational burden and rationale behind their strategies.

\item Based on the study, we distill the results into a list of suggestions from the perspective of patch authors, maintainers, and the community at large to improve the patch porting process.

\squishend

%\section{Background and Motivation}
\section{Background} \label{sec:background}

\begin{figure}[h]
\begin{center}
  \includegraphics[width=0.5\textwidth]{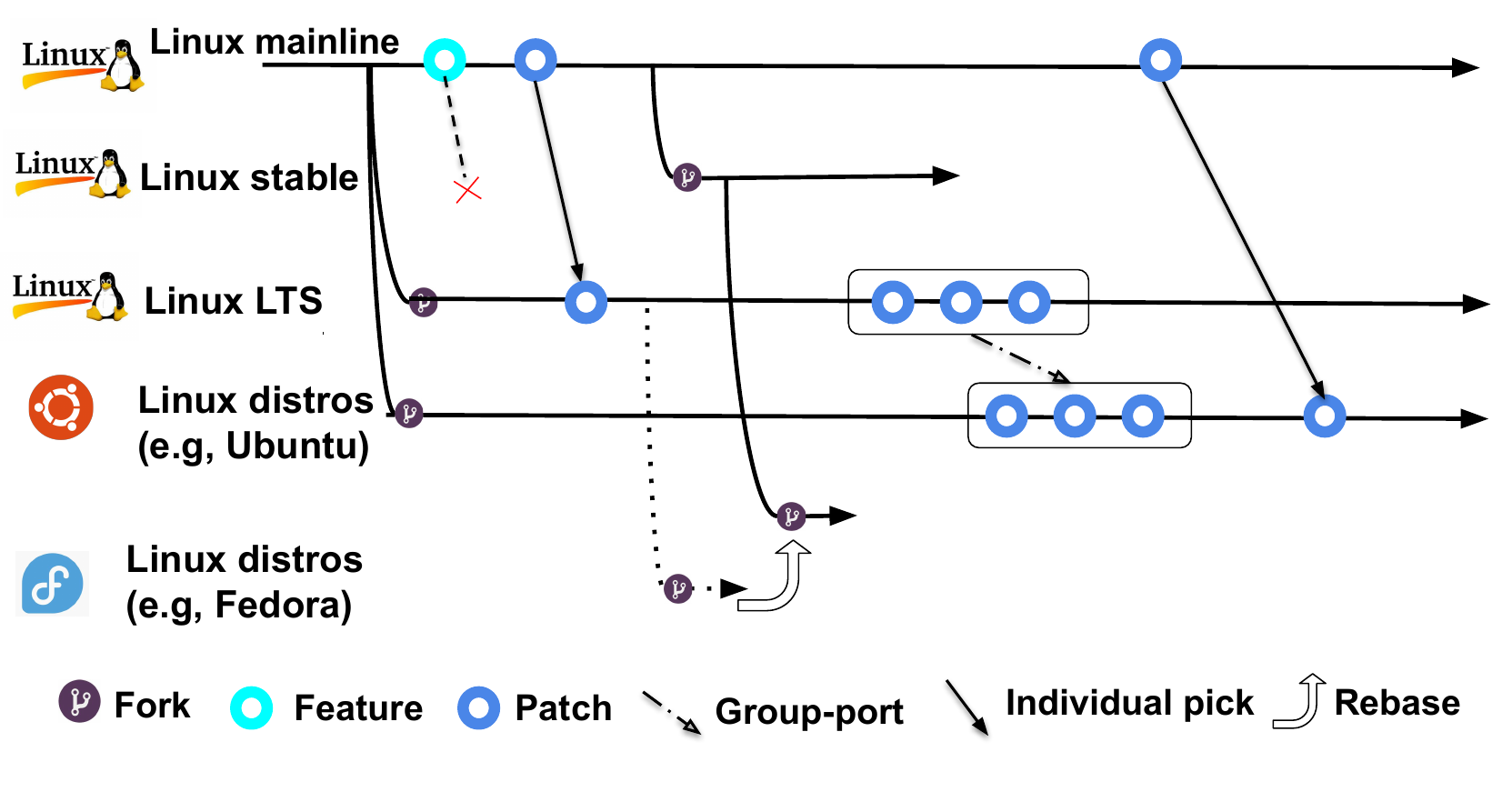}
\end{center}
\vspace{-.2in}
 \caption{Linux ecosystem and patch  process}
 \label{fig:Linux_eco}
 \vspace{-.2in}
\end{figure}

\PP{The Linux Ecosystem}
The Linux ecosystem's patch porting process is shown in ~\autoref{fig:Linux_eco}.
The trunk branch is the Linux mainline~\cite{kernelreleases}, which keeps integrating all the new features and bug fixes from a great number of kernel developers and maintainers.
The mainline branch is for development only and is not supposed to be used by end users directly.
Instead, whenever Linux mainline reaches a milestone, 
it is forked into either a stable or a long-term support (LTS) branch~\cite{kernelreleases}, e.g., v4.19.y is a LTS branch while v4.20.y is a stable branch.
In between the mainline releases, e.g., between v4.18 and v4.19, there are also release candidates such as v4.19-rc1 and v4.19-rc2 signifying the stages of the development process (e.g., most feature commits occur in rc1).
Linux stable and LTS branches provide stability by integrating only bug fixes from the mainline without adding any new features (according to some predefined rules~\cite{stable_kernel_rules}), that could introduce new bugs.
The main difference between Linux stable branches and LTS branches is their lifetime---stable branches usually are maintained for two to three months, whereas some Linux LTS branches have been maintained for six years~\cite{kernelreleases}.
Linux stable/LTS branches usually release a minor version (e.g., 4.15.3) once a week or once every two weeks~\cite{kernelreleases}.
Both the mainline and stable/LTS branches are maintained by the Linux community.

Outside of the branches maintained by the Linux community, we have branches maintained by Linux distributions, such as Ubuntu and Fedora.
They are what the end-users will be using in practice.
These distribution branches are typically a fork from one of the stable or LTS branches and typically include their own customizations (\eg better or additional hardware support).
Distributions can also fork from other distributions.
For example, the Linux Mint distribution uses the same kernel as Ubuntu.
As such branches are separately maintained, the burden of tracking and applying patches from stable/LTS or even mainline in a timely fashion is on distributions' maintainers.

\emph{Upstream and downstream:}
Depending on the forking and patch porting relationships, we have upstream and downstream branches.
The Linux mainline is often referred to as the {\em upstream}.
Linux distributions are usually referred to as {\em downstream}.
Of course, the relationship is also relative.
As shown in ~\autoref{fig:Linux_eco}, Linux stable/LTS branches are upstream with respect to distributions and downstream with respect to the mainline.

\PP{Patch Porting Practices.} 
In the Linux ecosystem, patches should be applied to the mainline first~\cite{patch-apply-process},
which are then ported to downstream kernels.
As mentioned, Linux stable/LTS branches port patches exclusively from Linux mainline.
For Linux distributions, there are four basic patch porting primitives employed by downstream kernel branches (one or more may be applied in a given downstream kernel branch).

\textit{Group-port:} For branches that are directly forked from stable/LTS branches, they port patches primarily from the corresponding upstream branch it is forked from.
Because patches from stable/LTS branches are usually ported in groups (\eg via \texttt{git merge}), we call this patch porting practice \emph{group-port}. 

\textit{Individual pick:}
Distribution maintainers can also choose to bypass the immediate LTS/stable branches and directly port patches from the mainline, when those patches are deemed urgent.
Because such patches are typically incorporated individually (\eg via the \texttt{git cherry-pick} command), we refer to this practice as \emph{picking}. It is worth noting that distributions may also directly pick feature commits from Linux mainline (e.g., for supporting new hardware devices).
We found that the picking is usually driven by each vendor's own bug tracking system, e.g., the launchpad system\cite{driving_force_case10,driving_force_case7,driving_force_case8, driving_force_case3} in Ubuntu, the OraBug system\cite{oracle_bug_tracking} in Oracle-Uek and the bugzilla system\cite{sle_bug_tracking_example} in SLE.

\textit{Minor version rebase:}
Different from selecting what patches to port, \emph{rebase} unconditionally changes the base itself to a given point in a target branch, e.g., latest mainline, stable or LTS branches, automatically inheriting all commits that were made to the target. A minor version rebase indicates that it rebases from one minor version to another (e.g., from 4.15.1 to 4.15.2). It is similar in effect to group-port except that it by default ports/inherits all commits between two minor versions (whereas group-port can and does omit some).

\textit{Major version rebase:}
Contrary to a minor version rebase, a major version rebase represents a rebase that goes from one major version to another (e.g., from 4.15.y stable to 4.16.y). This implies a significant change in the branch as there can be a large number of commits (patches or even feature commits) between the two major versions, all of which are inherited automatically after the rebase.

\PP{Fixes tag and Cc stable tag}
\phantomsection
\label{tags}
Mainline patches can be attached with two kinds of tags which make the patch porting easier for maintainers:

\squishlist
\item A fixes tag looks like the following: \texttt{Fixes: a9e38c3e01ad ("KVM: x86: Catch potential overrun in MCE setup")}.
It gives the commit (hash) that introduced the bug that is now patched. Such a tag is generated by the author(s) of the commit
typically after manually analyzing the historical Linux versions.
It is helpful for maintainers to determine if a kernel version is affected by the bug fixed by the patch.
In fact, it can also allow one to compute the bug lifetime, which was interestingly never used in previous measurement studies~\cite{alexopoulos2022long,li2017large}.
\item A Cc stable mailing list tag looks like the following: \path{Cc: stable@vger.kernel.org}.
This indicates that the author of the patch has determined that this commit satisfied the patch rules and should be considered for inclusion in stable/LTS branches. 
\squishend

\cut{
 \begin{figure}[t]
\begin{center}
  \includegraphics[width=0.5\textwidth]{figs/Motivating_example.pdf}
\end{center}
\vspace{-.2in}
 \caption{Patch propagation of CVE-2018-20961}
 \label{fig:motivating}
 \vspace{-.2in}
\end{figure}
}

\section{Dataset, Metrics, and Methodology}
\label{sec:data_method}

\PP{Measurement targets and dataset}
\cut{Besides the Linux mainline, we choose to measure six Linux LTS branches that are actively being maintained:
4.4 LTS, 4.9 LTS, 4.14 LTS, 4.19 LTS, 5.4 LTS, 5.10 LTS and 5.15 LTS. }
For Linux distributions, we choose to measure a few representative popular distributions according to a list maintained by LWN~\cite{LWNdistributions}
which has been tracking Linux distributions since 1999. 
Specifically, we study the following: four Ubuntu branches, three SUSE Linux Enterprise (SLE) branches, four Oracle Unbreakable Enterprise Kernel (UEK) branches, three Amazon Linux 2 branches, two Debian branches, three Android branches, one Fedora branch, and one Arch branch (both Fedora and Arch maintain a single kernel branch which keeps moving forward). 
In the case of Android, we picked three branches from Android-common~\cite{android_common_introduction}, which is part of the Android Open Source Project (AOSP),
further forked by downstream Android OEMs such as Samsung.
We exclude OEM kernels due to the fact that OEMs, such as Samsung, are known to not publicly release their git repositories, which are required for our study.
Note that there are other popular distributions such as Redhat and CentOS that we do not include. 
This is mostly because of the lack of public git repositories or insufficient data. For example, Redhat Enterprise Linux is not free. 
Alternatively, CentOS's role has been shifting over the past few years---initially being a downstream of Redhat, but now CentOS is becoming the upstream
of Redhat~\cite{centos_role}.  
In addition, we also measure six Linux LTS branches that are actively being maintained \cite{kernelreleases}:
4.4 LTS, 4.9 LTS, 4.14 LTS, 4.19 LTS, 5.4 LTS, 5.10 LTS and 5.15 LTS. 
We omit the Linux stable branches because the lifetime of stable branches is only two to three months, which is often less than the patch delays observed in LTS branches. Overall, our measurement targets include \textbf{{\em 584K}} commits. 

We list the patch porting strategies for all measured distribution branches in \autoref{table:branch_stategies}. Specifically, Arch and Fedora are the two distributions that stay on Linux stable/LTS branches only for a short period of time (usually for two to three months) and frequently rebase to new major release versions (e.g., from 4.14.y to 4.15). 
All other distributions choose to stay on a single base (i.e., a particular major release version) for an extended period of time (in years). While the distributions stay on a base, they choose a variety of patch porting practices such as minor version rebase, group-port, and picking from mainline. Most distributions stick to a consistent strategy. Ubuntu-18.04 is an exception 
where it first group-ports from Linux 4.15 stable, and then it group-ports from Linux 4.14 LTS and 4.19 LTS. 
SLE is also exceptional as it first group-ports from a stable release (instead of LTS), then it will directly pick commits from mainline after the stable branch ends (typically in two to three months).

\begin{table*}[]
\centering
\begin{tabular}{|l|c|c|c|c|c|c|}
\hline
                                & \multicolumn{1}{l|}{\begin{tabular}[c]{@{}l@{}}Major version \\ frequently rebase\end{tabular}} & \multicolumn{1}{l|}{\begin{tabular}[c]{@{}l@{}}Minor version\\ rebase on stable\end{tabular}} & \multicolumn{1}{l|}{\begin{tabular}[c]{@{}l@{}}Minor version \\ rebase on LTS\end{tabular}} & \multicolumn{1}{l|}{\begin{tabular}[c]{@{}l@{}}Group-port\\  from LTS\end{tabular}} & \multicolumn{1}{l|}{\begin{tabular}[c]{@{}l@{}}Group-port \\ from stable\end{tabular}} & \multicolumn{1}{l|}{\begin{tabular}[c]{@{}l@{}}Picking from \\ mainline\end{tabular}} \\ \hline
Arch                            & \ding{52}                                                                                               & \ding{52}                                                                                               & \ding{52}                                                                                            &                                                                                     &                                                                                        &                                                                                       \\ \hline
Fedora                          & \ding{52}                                                                                                &                                                                                               &                                                                                             & \ding{52}                                                                                     & \ding{52}                                                                                        &                                                                                       \\ \hline
Oracle Uek-4                    &                                                                                                 &                                                                                               &                                                                                             &                                                                                     & \ding{52}                                                                                       & \ding{52}                                                                                      \\ \hline
Oracle Uek-5 / 6 / 7            &                                                                                                 &                                                                                               &                                                                                             & \ding{52}                                                                                    &                                                                                        & \ding{52}                                                                                      \\ \hline
Ubuntu-16.04 / 20.04 / 22.04    &                                                                                                 &                                                                                               &                                                                                             & \ding{52}                                                                                    &                                                                                        & \ding{52}                                                                                     \\ \hline
Ubuntu-18.04                    &                                                                                                 &                                                                                               &                                                                                             &\ding{52}                                                                                    & \ding{52}                                                                                       &\ding{52}                                                                                      \\ \hline
SLE12-SP5 / 15- SP3 / 15-SP4    &                                                                                                 &                                                                                               &                                                                                             &                                                                                     & \ding{52}                                                                                       & \ding{52}                                                                                      \\ \hline
Amazon-4.14 / 5.4 / 5.10 &                                                                        &                                                                                               &             \ding{52}                                                                                 &                                                                                     &                                                                                        & \ding{52}                                                                                      \\ \hline
Debian-10 / 11 &                                                                        &                                                                                               &                                                                                           &                          \ding{52}                                                              &                                                                                        & \ding{52}                                                                                      \\ \hline
Android-10 / 12 / 13 &                                                                        &                                                                                               &                                                                                           &                          \ding{52}                                                              &                                                                                        & \ding{52}                                                                                      \\ \hline
\end{tabular}
\caption{Patch porting strategies for all measured distribution branches}
\label{table:branch_stategies}
\end{table*}

\PP{Measurement metrics}
We define the following three metrics to evaluate the patch porting performance:

\squishlist

\item \textit{Patch delay:} New Linux patches are officially released in Linux mainline first. So, we compute the patch delay for each patch applied to a downstream kernel branch as the time difference between when the patch appears in the mainline and when it appears in a downstream kernel branch. If the branch is a Linux distribution (e.g., Ubuntu), the patch delay is the end-to-end delay where the patch may be initially ported from mainline to an LTS branch first and then group-ported by the distribution.

\item \textit{Patch rate:} For each downstream kernel branch, we count the number of ported patches per day to compare the patching effort among distributions. The intuition is that it is challenging to obtain the ground truth in terms of the total number of patches that should be applied to a particular downstream kernel branch. However, approximately speaking, if a downstream kernel branch is well-maintained, it should periodically port enough patches. This metric will allow us to loosely compare the diligence of patch porting for various downstream kernel branches.

\item \textit{Bug inheritance ratio:} This metric is the fraction of bugs introduced into the mainline that are inherited by a downstream kernel branch. The intuition is that the mainline branch is the development branch that constantly introduces bugs (e.g., because of features commits or even bugs in patch commits). If a downstream kernel ports these bug-introducing commits from mainline, it will naturally inherit these bugs, which include a subset of mainline bugs (i.e., the ratio is always smaller than or equals 100\%).

\squishend

\PP{Methodology to compute patch delays}
Recall that all patches in the Linux ecosystem will first appear in mainline,
so to calculate the patch porting delay for each patch that appears in a downstream branch, 
we need to first determine which mainline commit it corresponds to.

\textit{Tracking patch origins for Linux stable/LTS branches:}
we observe that 99.3\% of all the git commits in LTS branches contain references to an upstream commit in the forms like
``\texttt{Upstream commit <hash>}'', ``\texttt{upstream <hash> commit}'' in the git commit messages.
This is supported by the official script~\cite{git_format_lts}  used by Linux LTS maintainers to port patches from mainline.

\textit{Tracking patch origins for Linux distributions:}
As mentioned earlier, distributions have three typical approaches to port patches: group-porting, picking, and rebasing.

For group-porting, we found two commonly used methods: \texttt{git merge} and the Ubuntu-specific method.

\squishlist
 
\item \texttt{git merge} ports all commits from a specific stable/LTS branch up to a certain commit specified in the command.  
It will generate one merge commit with the title that usually indicates the latest point that it caught up with.
For instance, \texttt{Merge tag 'v4.14.107'} indicates group-porting patches reached a specified point in the source branch, which is the version of v4.14.107.
In our measurement targets, Oracle-UEK and SLE use this method. 

\item We find that Ubuntu takes a different approach to apply group-ports of upstream patches in a single commit, without using \texttt{git merge}~\cite{KernelBugFixingUbuntu}.
We observe that such commits will contain links to the corresponding bug tracking pages on \texttt{launchpad.net} (\eg \cite{launchpad_example_page}),
which contain detailed information about what patches have been ported and from where.  

\squishend

Rebase is performed using \texttt{git rebase}, which changes the entire base (\ie the fork point) of a branch from one upstream commit to another (there is the major version rebase vs. minor version rebase as mentioned previously).
This also results in the porting of all commits between the old base and the new base.
It is worth noting that \texttt{git rebase} will not introduce a new commit in the log.
In fact, the log will essentially be a replica of the upstream to which it rebases. This makes it difficult to infer when the rebase occurs and compute the patch delays accordingly.
Nevertheless, distribution maintainers 
%of measurement targets 
always will add a followup commit right after the rebase to assign a unique tag and indicate the rebase has been completed.
This commit (\eg with a title of \texttt{kernel-5.10.23-200.fc33}) usually indicates the mainline/stable/LTS version of the distribution from which it is rebased.
This allows us to determine when rebase occurs, from which we can then compute the patch porting delay.

When Linux distributions pick patches directly from the mainline,
each commit message will contain text referencing the corresponding commit's hash in mainline, such as ``\texttt{cherry picked from <hash>}'' and ``\texttt{back-ported from <hash>}''.
Because such references appear in different areas from those that appear in LTS commit messages,
this allows us to distinguish picked patches from group-ported ones.

Another issue is that we find some distributions pick features commits in addition to patches from mainline,
which we need to exclude when computing their patch delay. 
For Ubuntu, we find that feature commits picked from mainline are managed by its own bug tracking system on launchpad.com, and labeled as ``[Feature]'' on the web page.
For SLE, they have a separate tracking system for feature requests on \url{jiri.suse.com} and attach identifiers that start with ``jscSLE'' to denote that it is a feature commit. 
For Oracle-Uek, all commits picked directly from Linux mainline are labeled ``OraBug'' id in commit messages, indicating they are bug fixes. For other distributions, such as Fedora, according to our observations and direct communications with distribution maintainers, they only pick patches (and no feature commits) from Linux mainline.

\PP{Methodology to compute patch rate}
For a given downstream kernel branch, we divide the total number of patches that appear in the downstream branch (e.g., no matter if they are group-ported, picked, or inherited due to rebase) by the time duration starting with the first patch appearance date and ending with the last patch appearance date.
In our measurements, we typically choose the duration starting with the very first commit since the creation of a downstream branch, and ending with the last commit we observe at the time of writing. If the branch has ended its end of life, we effective computes the average patch rate over its entire lifetime.

\begin{figure*}
\centering
% \hspace{-.1in}
    \begin{minipage}[t]{.3\linewidth} % A minipage that
        \centering
        \includegraphics[scale=0.155]{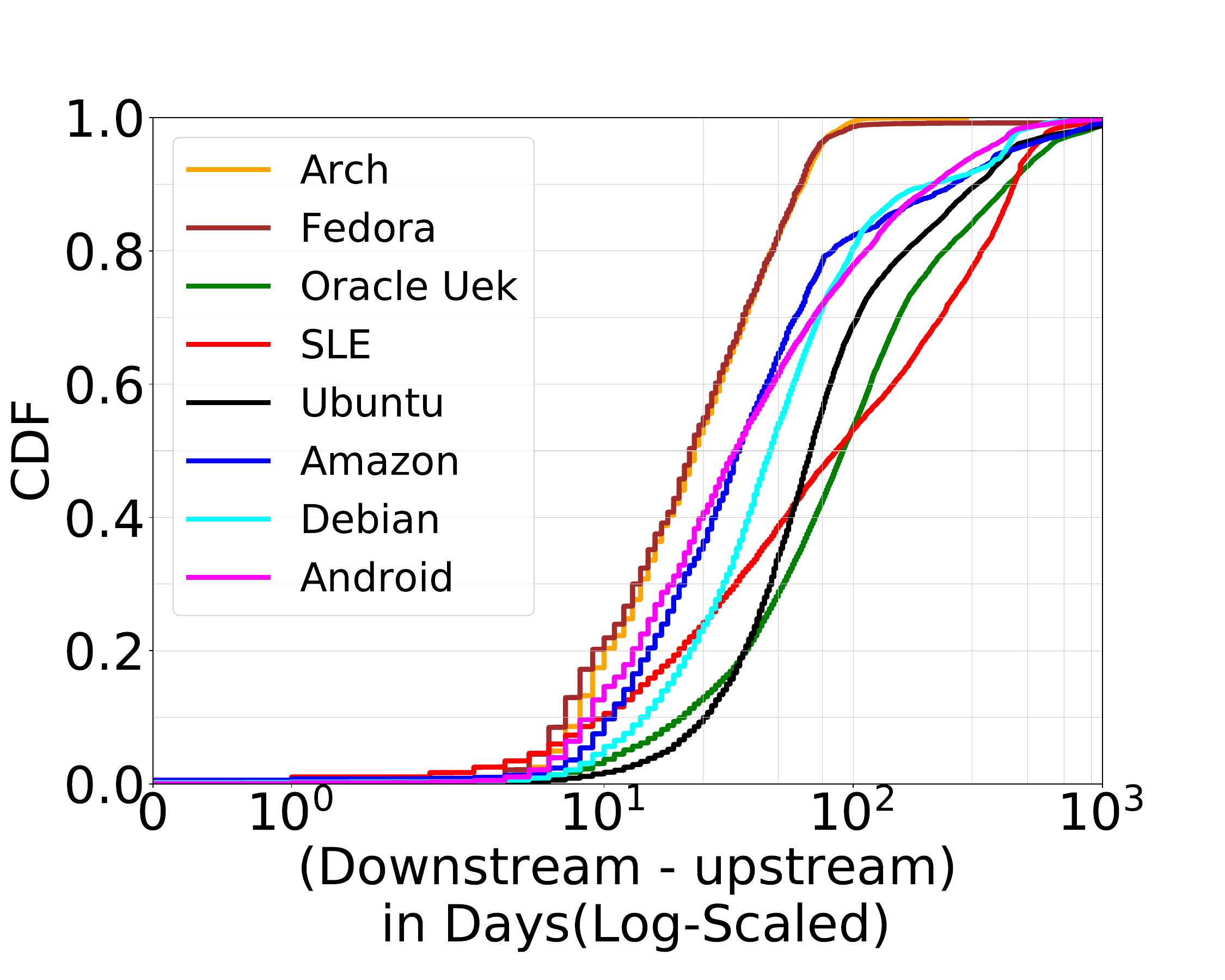}
         \vspace{-.2in}
        \caption{End-to-end delay for all patches in Linux distributions } 
        % \vspace{-.1in}
        \label{fig:delay_alldistributions}
        \vspace{-.1in}
    \end{minipage}
    \begin{minipage}[t]{.3\linewidth} % A minipage that
        \centering
        \includegraphics[scale=0.155]{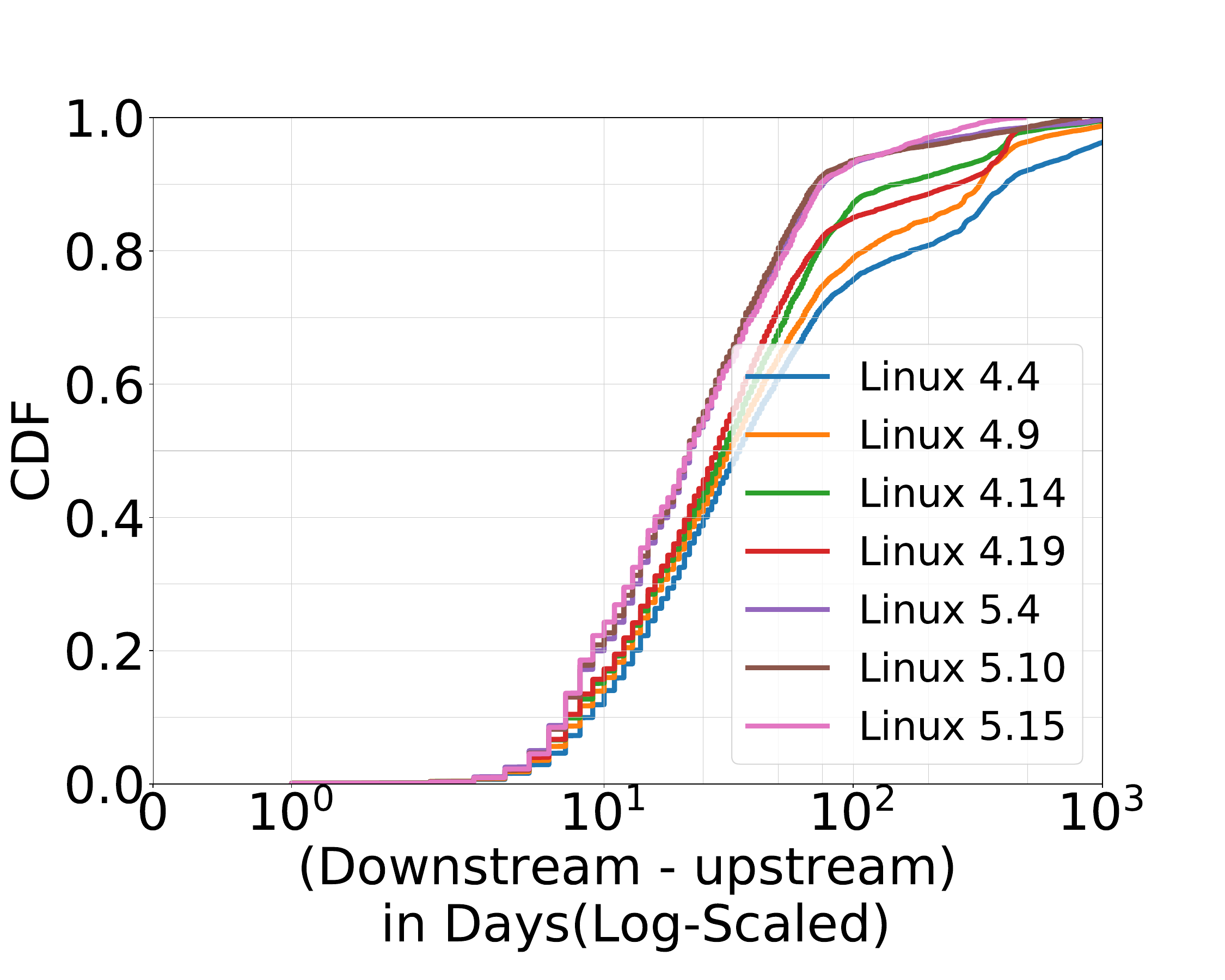}
        \vspace{-.2in}
        \caption{End-to-end delay for all patches in Linux LTS branches} 
        % \vspace{-.1in}
        \label{fig:LTS_delay}
        \vspace{-.1in}
    \end{minipage}
        \begin{minipage}[b]{0.3\linewidth} % A minipage that
        \centering
        \includegraphics[scale=0.155]{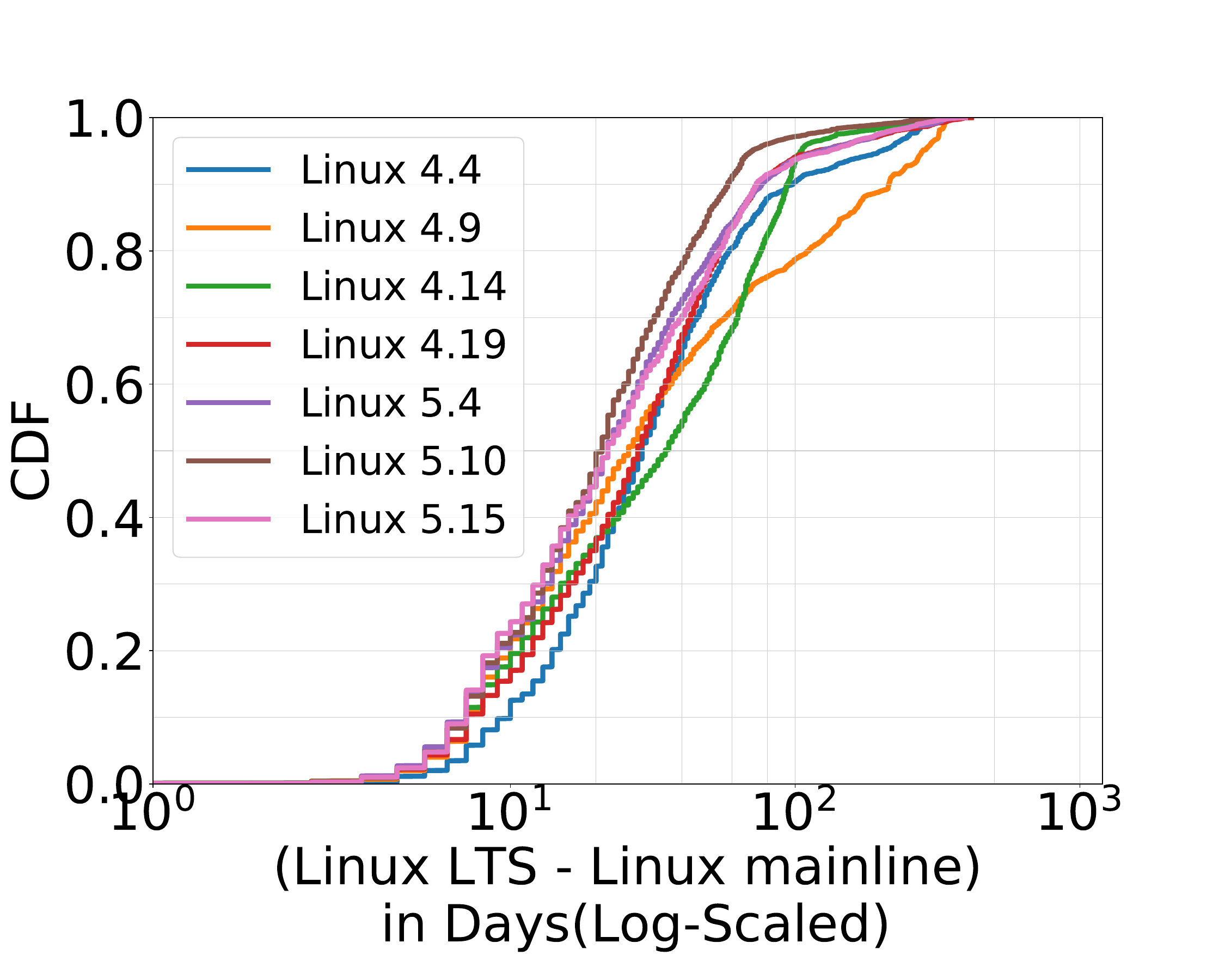}
        \vspace{-.2in}
        \caption{First year delay since the creation of Linux LTS branches}   
        \label{fig:LTS_first_year}
        \vspace{-.2in}
    \end{minipage}
\end{figure*}

\PP{Methodology to compute bug inheritance ratio}
We define a mainline bug as a commit (either a feature or patch commit) that requires a later patch.
As mentioned in the previous section, fixes tags in a patch indicate which previous commit is the buggy commit,
so we use fixes tags to identify mainline bugs. 
To validate the accuracy of the fixes tag, we randomly selected 50 buggy commits based on the fixes tags that have reproducers, obtained from syzbot~\cite{syzbot}, a public continuous fuzzer for Linux kernels. We then automatically executed these reproducers against Ubuntu branches that contained the buggy commits (but unfixed yet), and found that 45 of them were able to successfully trigger the bug. This indicates that fixes tags are generally reliable.
We consider a bug inherited by a downstream kernel branch if the buggy mainline commit ported to the downstream branch,
through either group-port, picking, or rebase.
Note that we consider bugs inherited only after the downstream kernel branch is created.
In other words, a downstream branch has zero bugs inherited at the time of its creation.
The bug inheritance ratio is then computed by the total number of bugs inherited by a downstream branch divided by the total number of bugs in mainline, in a given time period.
Again, in our measurements, we often choose the time window to be the lifetime of a downstream kernel branch to understand its average performance.
We note that even though there is an underestimate of mainline bug count as not all patches have fixes tags,
the ratio is a relative metric which makes it possible to compare the performance of different downstream kernel branches.

% \section{Rebase vs Non-rebase}
\section{Strategy of Frequently Changing Bases}

\begin{table*}[!h]
\centering
\resizebox{\linewidth}{!}{
\begin{tabular}{|r|rllllll|llllll|}
\hline
\multicolumn{1}{|l|}{}     & \multicolumn{7}{c|}{By all ported patches}                                                                                                                                                                              & \multicolumn{6}{c|}{By picking from Linux mainline}                                                                                                                              \\ \hline
\multicolumn{1}{|l|}{base} & \multicolumn{1}{l|}{Linux LTS}   & \multicolumn{1}{l|}{Oracle-Uek}  & \multicolumn{1}{l|}{Ubuntu}      & \multicolumn{1}{l|}{Amazon}     & \multicolumn{1}{l|}{SLE}         & \multicolumn{1}{l|}{Debian}     & Android & \multicolumn{1}{l|}{Oracle-Uek} & \multicolumn{1}{l|}{Ubuntu}    & \multicolumn{1}{l|}{Amazon}    & \multicolumn{1}{l|}{SLE}         & \multicolumn{1}{l|}{Debian}     & Android \\ \hline
4.1                        & \multicolumn{1}{r|}{}            & \multicolumn{1}{r|}{1.7\%/3.3}   & \multicolumn{1}{l|}{}            & \multicolumn{1}{l|}{}           & \multicolumn{1}{l|}{}            & \multicolumn{1}{l|}{}           &         & \multicolumn{1}{r|}{1.6\%/2.8}  & \multicolumn{1}{l|}{}          & \multicolumn{1}{l|}{}          & \multicolumn{1}{l|}{}            & \multicolumn{1}{l|}{}           &         \\ \hline
4.4                        & \multicolumn{1}{r|}{2.7\%/8.4}   & \multicolumn{1}{l|}{}            & \multicolumn{1}{r|}{3.5\%/10.1}  & \multicolumn{1}{l|}{}           & \multicolumn{1}{l|}{}            & \multicolumn{1}{l|}{}           &         & \multicolumn{1}{l|}{}           & \multicolumn{1}{r|}{1.0\%/1.5} & \multicolumn{1}{l|}{}          & \multicolumn{1}{l|}{}            & \multicolumn{1}{l|}{}           &         \\ \hline
4.9                        & \multicolumn{1}{r|}{3.4\%/10.4}  & \multicolumn{1}{l|}{}            & \multicolumn{1}{l|}{}            & \multicolumn{1}{l|}{}           & \multicolumn{1}{l|}{}            & \multicolumn{1}{l|}{}           &         & \multicolumn{1}{l|}{}           & \multicolumn{1}{l|}{}          & \multicolumn{1}{l|}{}          & \multicolumn{1}{l|}{}            & \multicolumn{1}{l|}{}           &         \\ \hline
4.12                       & \multicolumn{1}{r|}{}            & \multicolumn{1}{l|}{}            & \multicolumn{1}{l|}{}            & \multicolumn{1}{l|}{}           & \multicolumn{1}{r|}{17.6\%/35.1} & \multicolumn{1}{l|}{}           &         & \multicolumn{1}{l|}{}           & \multicolumn{1}{l|}{}          & \multicolumn{1}{l|}{}          & \multicolumn{1}{r|}{17.4\%/34.7} & \multicolumn{1}{l|}{}           &         \\ \hline
4.14                       & \multicolumn{1}{r|}{4.5\%/13.3}   & \multicolumn{1}{r|}{6.7\%/15.9}    & \multicolumn{1}{l|}{}            & \multicolumn{1}{r|}{5.1\%/14.0}   & \multicolumn{1}{l|}{}            & \multicolumn{1}{l|}{}           &         & \multicolumn{1}{r|}{3.1\%/4.2}  & \multicolumn{1}{l|}{}          & \multicolumn{1}{r|}{0.3\%/0.4} & \multicolumn{1}{l|}{}            & \multicolumn{1}{l|}{}           &         \\ \hline
4.15                       & \multicolumn{1}{r|}{}            & \multicolumn{1}{l|}{}            & \multicolumn{1}{r|}{5.1\%/14.7}  & \multicolumn{1}{l|}{}           & \multicolumn{1}{l|}{}            & \multicolumn{1}{l|}{}           &         & \multicolumn{1}{l|}{}           & \multicolumn{1}{r|}{1.3\%/1.8} & \multicolumn{1}{l|}{}          & \multicolumn{1}{l|}{}            & \multicolumn{1}{l|}{}           &         \\ \hline
4.19                       & \multicolumn{1}{r|}{5.1\%/16.1}  & \multicolumn{1}{l|}{}            & \multicolumn{1}{l|}{}            & \multicolumn{1}{l|}{}           & \multicolumn{1}{l|}{}            & \multicolumn{1}{l|}{3.2\%/10.1} & 5.2\%/16.05  & \multicolumn{1}{l|}{}           & \multicolumn{1}{l|}{}          & \multicolumn{1}{l|}{}          & \multicolumn{1}{l|}{}            & \multicolumn{1}{l|}{0.7\%/0.2}  & 0.1\%/0.1    \\ \hline
5.3                        & \multicolumn{1}{r|}{}            & \multicolumn{1}{l|}{}            & \multicolumn{1}{l|}{}            & \multicolumn{1}{l|}{}           & \multicolumn{1}{r|}{14.3\%/35.0} & \multicolumn{1}{l|}{}           &         & \multicolumn{1}{l|}{}           & \multicolumn{1}{l|}{}          & \multicolumn{1}{l|}{}          & \multicolumn{1}{r|}{13.3\%/33.0} & \multicolumn{1}{l|}{}           &         \\ \hline
5.4                        & \multicolumn{1}{r|}{6.3\%/19.9}  & \multicolumn{1}{r|}{11.4\%/24.7} & \multicolumn{1}{r|}{6.6\%/18.9}  & \multicolumn{1}{r|}{6.9\%/19.0} & \multicolumn{1}{l|}{}            & \multicolumn{1}{l|}{}           & 5.8\%/18.5   & \multicolumn{1}{r|}{5.4\%/6.3}  & \multicolumn{1}{r|}{1\%/1.3}   & \multicolumn{1}{r|}{0.3\%/0.3} & \multicolumn{1}{l|}{}            & \multicolumn{1}{l|}{}           & 1.1\%/1.2    \\ \hline
5.10                       & \multicolumn{1}{r|}{8.8\%/24.8}  & \multicolumn{1}{l|}{}            & \multicolumn{1}{l|}{}            & \multicolumn{1}{r|}{10.5\%/24.7}  & \multicolumn{1}{l|}{}            & \multicolumn{1}{l|}{6.7\%/17.3} &   9.1\%/23.0      & \multicolumn{1}{l|}{}           & \multicolumn{1}{l|}{}          & \multicolumn{1}{r|}{0.5\%/0.6} & \multicolumn{1}{l|}{}            & \multicolumn{1}{l|}{0.2\%/0.21} &  1.4\%/2.1        \\ \hline
5.14                       & \multicolumn{1}{r|}{}            & \multicolumn{1}{l|}{}            & \multicolumn{1}{l|}{}            & \multicolumn{1}{l|}{}           & \multicolumn{1}{r|}{15.4\%/49.0} & \multicolumn{1}{l|}{}           &         & \multicolumn{1}{l|}{}           & \multicolumn{1}{l|}{}          & \multicolumn{1}{l|}{}          & \multicolumn{1}{r|}{9.7\%/39.8}  & \multicolumn{1}{l|}{}           &         \\ \hline
5.15                       & \multicolumn{1}{r|}{13.1\%/30.8} & \multicolumn{1}{r|}{13.3\%/30.3} & \multicolumn{1}{r|}{12.3\%/31.0} & \multicolumn{1}{r|}{}           & \multicolumn{1}{l|}{}            & \multicolumn{1}{l|}{}           &         & \multicolumn{1}{r|}{1.4\%/1.5}  & \multicolumn{1}{r|}{2.7\%/2.8} & \multicolumn{1}{r|}{}          & \multicolumn{1}{l|}{}            & \multicolumn{1}{l|}{}           &         \\ \hline
\end{tabular}
}
\caption{Bug inheritance ratio / patch rate for Linux LTS branches and six distributions that stay on a single base}
\label{table:bug_inherritance_patch_rates}
\end{table*}

As mentioned in ~\autoref{table:branch_stategies}, Arch and Fedora frequently change their bases from one major version to another (every two to three months). This is intuitively a very different strategy compared to staying on a single base for an extended period of time (i.e., years).
In this section, we focus on the pros and cons of the frequent base changing strategy with respect to the metrics we defined.

In terms of patch delay, frequently changing the base is a good strategy because it automatically inherits all commits from the upstream, allowing the downstream kernel branch to keep up with the latest patches.
Indeed, we can see in ~\autoref{fig:delay_alldistributions} that Arch and Fedora have the lowest patch delay overall compared to other distributions. More specifically, almost all patches in Arch and Fedora are ported within 100 days. This is because all mainline patches are guaranteed to end up in an Arch or Fedora branch every time they rebase, which happens every two to three months.
Of course, while they stay on the same base, they can port patches earlier as well, e.g., Arch would perform minor version rebases and Fedora would perform group-ports from stable/LTS.

Unfortunately, when switching bases from one major version to another (i.e., major version rebase), branches also inherit new bugs as well as patches from the mainline.
For example, when a rebase occurs from v4.14.y to the beginning of v4.15, there are a large number of commits
between v4.14 and v4.15 on mainline that are inherited (e.g., due to the development of v4.15-rc1, v4.15-rc2).
All of these commits will be inherited automatically, which means the bug inheritance ratio is always 1.
This can be a significant downside leading to stability and security risks.
It is worth noting that minor version rebase is less of a concern because all commits in stable and LTS are patches by design.
In contrast, for distributions that choose to stay on a single base, their bug inheritance ratio is much lower, as shown in ~\autoref{table:bug_inherritance_patch_rates}. We will go into more detail in the next section regarding these measurements.

We suspect that the decision of frequently changing the base is driven by economics. Specifically, Arch and Fedora are community-backed distributions~\cite{community_commercial,fedora_community} which have limited resources invested to maintain a stable branch on their own. 
In contrast, the other distributions are considered commercial distributions~\cite{community_commercial,ubuntu_commerical}. For example, Amazon Linux 2 distributions are critical for AWS users where stability and security are critical\cite{amazon_linux}. As a result, these distributions prefer to stay on a single major version (e.g., v.4.14.y) for an extended period of time.

\section{Strategy of Staying on a Single Base}

Given that frequently changing bases 
%is not ideal as it 
inherits too many bugs from the mainline,
we now set that strategy aside and focus on the strategy of staying on a single base. 
Ideally, for this strategy, we would like to see that patches are ported as fast as possible,
and none of the necessary patches are missed.
We aim to analyze how far the downstream branches that follow this strategy, namely Linux LTS, Oracle-UeK, Ubuntu, SLE, Debain, Android and Amazon, are away from the ideal.

In particular, we will provide objective results with regard to individual metrics first 
and discuss unique behaviors in the patch porting practices for Linux distributions.
In addition, we will analyze the factors that influence the results which will help set the stage for future improvements.

\subsection{Patch delay}

According to the methods described in the previous section, we can measure the patch porting delays from Linux mainline to various downstream kernel branches, including Linux LTS branches and Linux distributions. ~\autoref{fig:LTS_delay} plots the patch delay between Linux mainline and Linux LTS branches and ~\autoref{fig:LTS_first_year} plots their delay for only the first year since the creation of Linux LTS branches, which shows that the three latest Linux LTS branches (i.e., Linux v5.4, v5.10 and v5.15) have noticeably smaller delays overall compared to older Linux LTS branches, indicating improvements over time. We do note that there are still about 5\% of the patches are delayed for more than 100 days even in the latest kernel branches. 
For older LTS branches, the performance is significantly worse, with 15\% to 25\% of patches are delayed more than 100 days.

Next, we plot the delay for Linux distributions, as shown in the ~\autoref{fig:delay_alldistributions}. 
Interestingly, unlike the similar performance between Arch and Fedora, Oracle-Uek, SLE, Ubuntu, Android, Debian and Amazon exhibit substantially different delay profiles. Amazon and Android have the lower delay compared with the four distributions, with about 20\% of patches delayed for 100 days or more. The worst delay goes to SLE, with about 20\% of patches ported after 300+ days. 
Note that it is expected that the patch delays for distributions are generally larger than those of LTS branches, because there is always an extra delay for a patch to be ported from LTS to distribution. 

\begin{figure*}
    \centering
    \begin{minipage}[t]{.3\linewidth} % A minipage that
        \centering
        \includegraphics[scale=0.155]{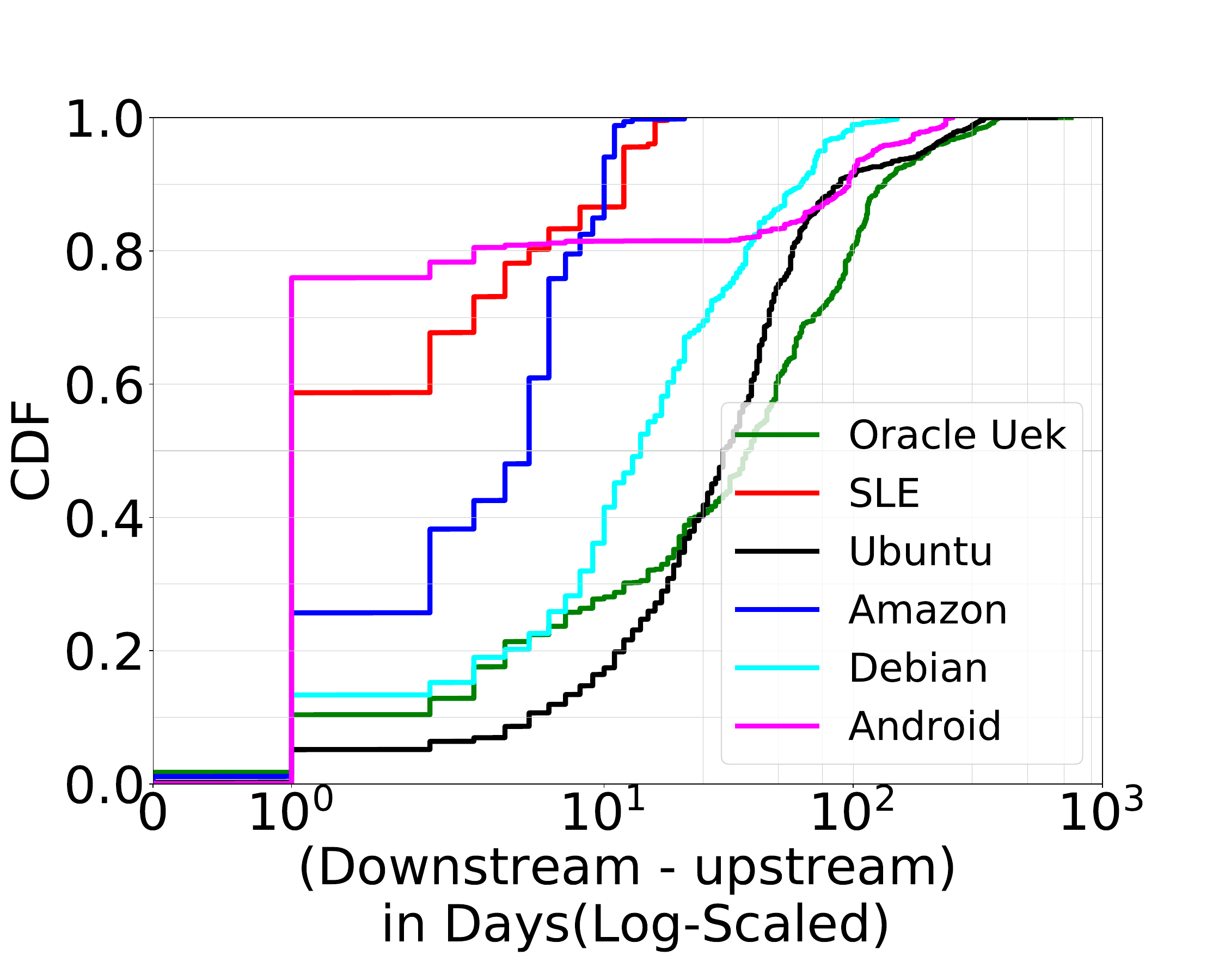}
       % \vspace{-.2in}
        \caption{Delay for Linux distribution commits whose sources are LTS}   
          \vspace{-.2in}
        \label{fig:groupport_delay}
        % \vspace{-.1in}
    \end{minipage}
     \begin{minipage}[t]{.3\linewidth} % A minipage that
        \centering
        \includegraphics[scale=0.155]{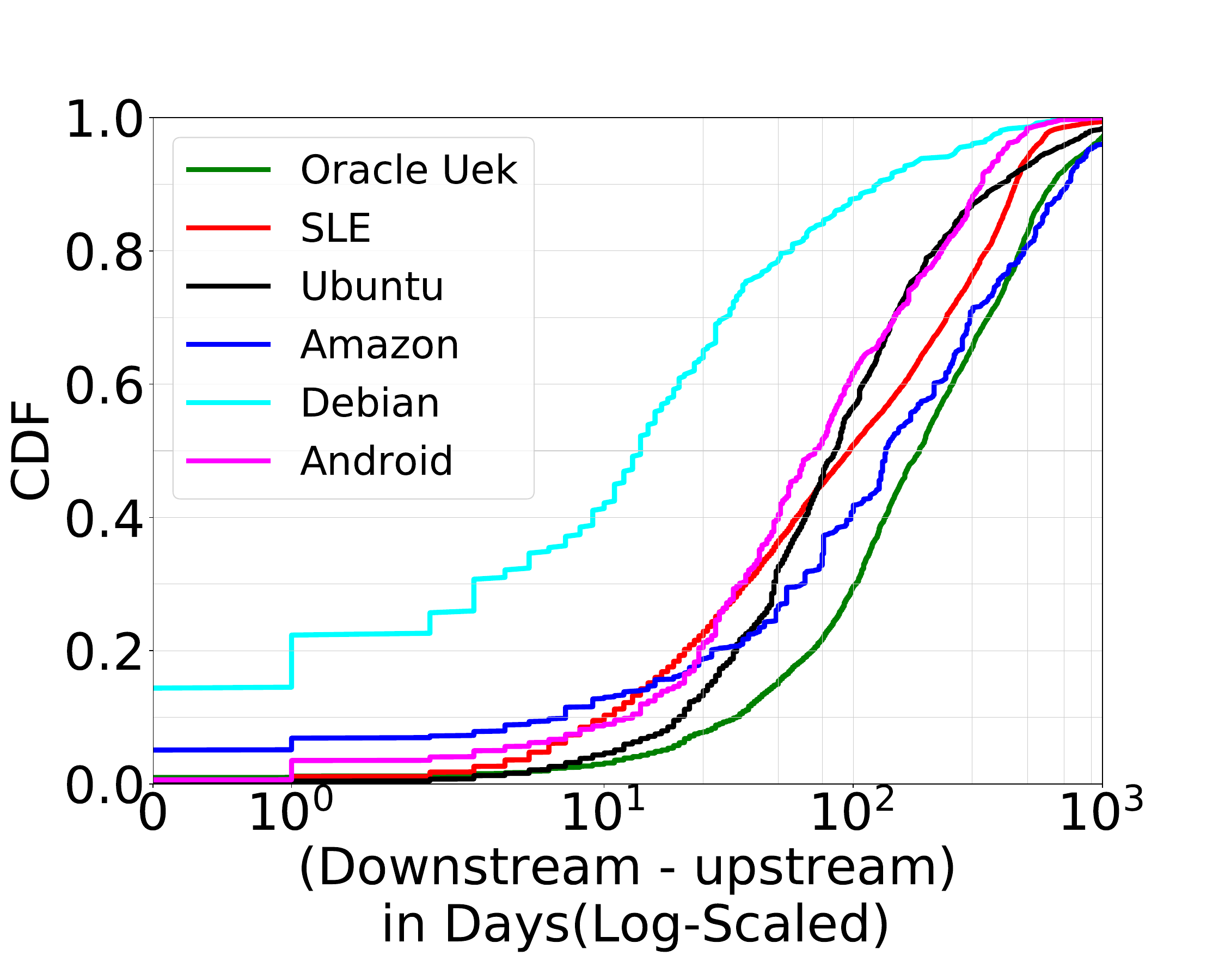}
         % \vspace{-.2in}
        \caption{Delay for Linux distribution commits whose sources are mainline}
        \label{fig:pick_delay}
     \end{minipage}
        \begin{minipage}[t]{.3\linewidth} % A minipage that
        % \begin{center}
        \centering
        \includegraphics[scale=0.155]{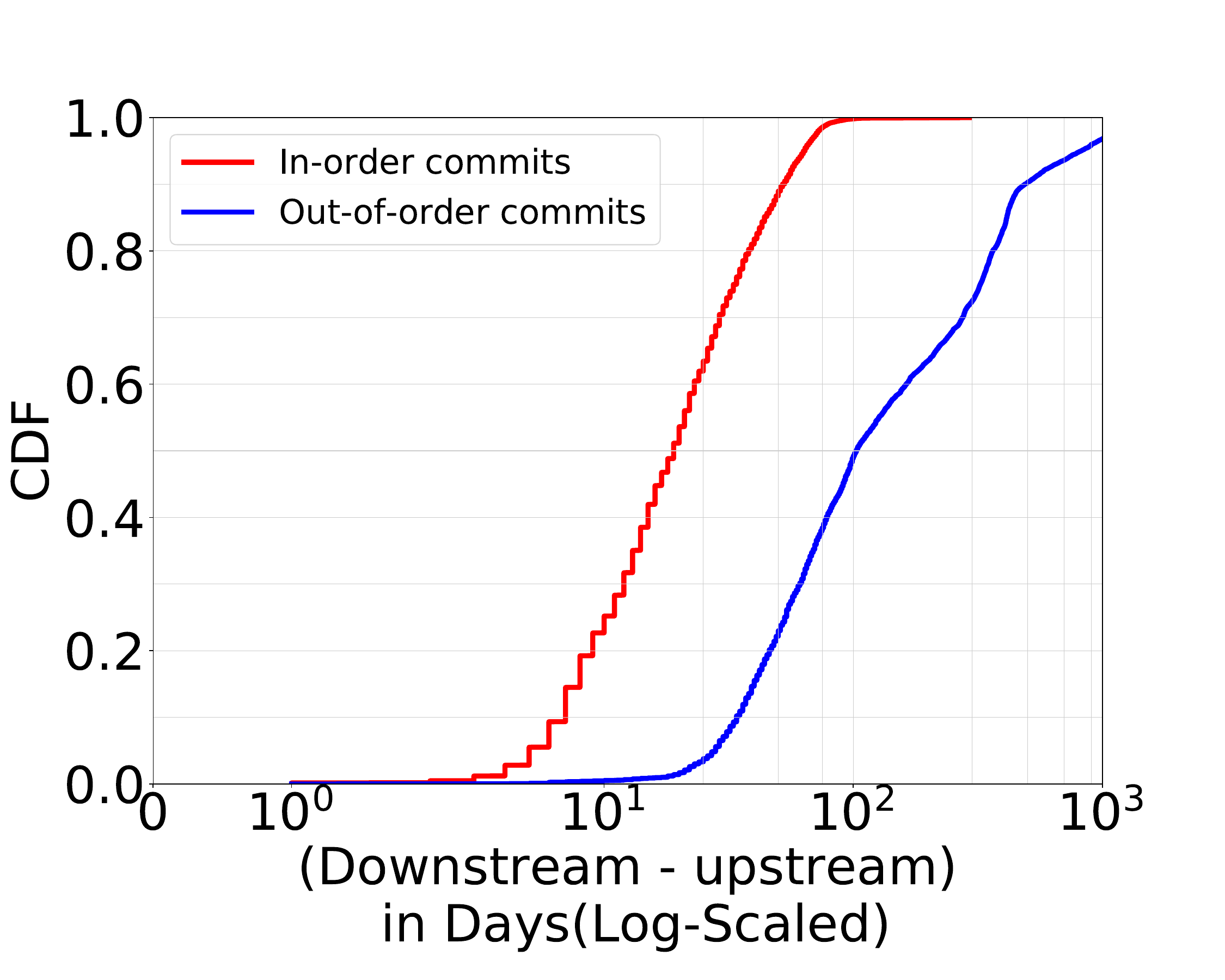}
        % \vspace{-.2in}
        \caption{The comparison of in-order and out-of-order commits} 
        % \vspace{-.2in}
        \label{fig:in_out_order_4_19}
    \end{minipage}
\end{figure*}

Finally, we breakdown the patch delay for Linux distributions. There are two patch porting routes for distributions: (1) Linux mainline to Linux LTS branches, and then to Linux distributions; (2) Linux mainline to Linux distributions directly (through picking).  ~\autoref{fig:LTS_delay} already shows the patch delay between Linux mainline and Linux LTS branches. 
The patch delay between Linux LTS and Linux distributions is shown in ~\autoref{fig:groupport_delay}. It clearly shows Android is much better than other distributions on the delay for group-porting. About 78\% of patches in Android are ported within one day.  However, there is a long tail with much longer delays compared to Amazon and SLE.
One possible reason is that the main maintainer of Android is also the main maintainer of Linux LTS branches~\cite{android_common_repo, linux_stable_tree}. 
Second to Android, all of the patches in Amazon and SLE are group-ported from LTS within 25 days. However, about 20\% of patches in Oracle-Uek are ported 100+ days after they appear in Linux LTS. ~\autoref{fig:pick_delay} plots the patch delay for picked patches with Linux mainline. Except Debian, all of the picked patches are delayed much longer than group-ported patches: 40\% - 70\% of picked patches are delayed 100+ days.

\cut{
that Amazon and SLE are much better than Ubuntu and Oracle-Uek. All of the patches in Amazon and SLE are group-ported from LTS within 25 days. However, about 24\% of patches in Oracle-Uek are ported 100+ days after they appear in Linux LTS. ~\autoref{fig:pick_delay} plots the patch delay for picked patches with Linux mainline. Different from group-porting from LTS, Amazon is the worst, while Ubuntu is the best. All of the picked patches are delayed much longer than group-ported patches: 40\% - 70\% of picked patches are delayed 100+ days.
}

% \begin{figure*}
%     \centering
%     \begin{minipage}[t]{.3\linewidth} % A minipage that
%         \centering
%         \includegraphics[scale=0.44]{figs/portDelay_bionic.pdf}
%          \vspace{-.1in}
%         \caption{The delay changes for Ubuntu-18.04 over the time} 
%         % \vspace{.1in}
%         \label{fig:portDelay_bionic}
%     \end{minipage}
% \vspace{-.2in}
% \end{figure*}

\subsection{Patch rate}

Patch rate is another critical facet of the patching porting performance. It is possible that some branch
has small patch delays but it ports \emph{only a small number of patches}. In other words, there may be many patches missing (or can be viewed as having an infinite delay). 

We show the results in ~\autoref{table:bug_inherritance_patch_rates}. 
Note that we compute the patch rate for ``all patches'' as well as ``patches picked from mainline''. This is because 
the picked patches from mainline can be viewed as extra work performed by distribution maintainers and worth highlighting separately.

There are several notable points from the table: 

1)  There is a general trend that newer branches (i.e., on newer major versions) have higher patch rate.
This is expected as older branches (e.g., v4.4) diverge more and more from mainline as time goes by,
and thus fewer patches are applicable. This is true for LTS branches and as well as most distributions (except for Oracle-Uek which is an outlier).

2) SLE has the highest patch rate among the six distributions --- much higher than distributions that are based on a similar kernel major version.
For example, SLE (v4.12) has 35.1 patches ported per day compared to Oracle-Uek (v4.14)'s 15.9 and Amazon's 14. 

3) Regarding the patch rate for picked patches only, we can see that SLE is also the highest. In fact, it picks more patches
per day compared to LTS. For example, for SLE (v5.3), its patch rate is 35 per day whereas LTS (v5.4) has a patch rate of 19.9 per day. Note that SLE's patch porting strategy is that it does group-porting initially from a stable branch and switches to exclusively picking from mainline after the stable branch reaches end-of-life. Effectively, SLE can be viewed as maintaining its own LTS branch.

4) Debian has the lowest patch rate among all distributions. For example, Debian (v5.10) has an overall patch rate of 17.3 compared to LTS (v5.10)'s 24.8 and Amazon (v5.10)'s 24.7. We will explain the reason behind this in \S\ref{debian}.

\subsection{Bug inheritance ratio}

The ideal patch porting does not only require that distributions port patches, but also that there should be no new bugs introduced during the process. We list the bug inheritance ratio in ~\autoref{table:bug_inherritance_patch_rates} as well.

We summarize a few notable points:

1) There is a general trend that newer branches have higher bug inheritance ratio. 
This is expected as we can see generally high bug inheritance ratios correlate with higher patch rate. 
Assuming there are always buggy patches that introduce bugs themselves, the more patches we pick, the more likely we will inherit those bugs.
\cut{
2) The most recent v5.15 major version has a much higher bug inheritance ratio of 13.1\% for LTS, much higher than the previous versions. For example, LTS v5.10 had a similar patch rate (only marginally smaller), its bug inheritance ratio is only 8.8%.
This is likely because v5.15 is relatively new and not all buggy commits are identified (smaller buggy commit base available for inheritance). Furthermore, because v5.15 is closest to mainline, a larger fraction of mainline patches will be applicable and ported. Together, this makes it more likely to inherit the buggy commits. 
}

2) The bug inheritance ratio is much higher in SLE branches than others, mainly contributed by the extensive picking from Linux mainline. This is likely again because of the substantially higher patch rate.

\subsection{Unique patch porting behaviors of Distributions}

As demonstrated above, six distributions exhibit varying results in three metrics, even though all of them choose to stay on the single base for a long time. This is due to various other unique (and sometimes subtle) behaviors in the patch porting practices.

\vspace{0.02in}
\noindent\textbf{SLE} 
effectively maintains its own LTS branches, as previously mentioned, 
by following short-lived Linux stable branches and picking patches from the mainline directly.
SLE appears determined to maintain high quality branches by aggressively selecting patches from the mainline. In fact, it selects significantly more patches compared to traditional LTS branches, showcasing a strong commitment in maintaining a reliable kernel.
As a result, its patch rate is high (in fact, the highest), it the bug inheritance ratio is also high (an expected tradeoff). 
However, even though it ports many patches, its patch delays are larger than most other distributions, especially for patch delays of 100+ days, likely due to the cost of picking more patches.

\vspace{0.02in}
\noindent\textbf{Oracle-Uek and Ubuntu}
overall employ a very similar strategy that involves grouping-porting from Linux LTS and picking from the mainline, and has comparable patch rate and bug inheritance ratio.
However, we have noticed significant periods of inactivity (i.e., idle periods), during which no group-porting takes place.
This is the key reason behind the noticeably worse group-port delay and end-to-end delay compared to other distributions. We will discuss this phenomenon more later in \S\ref{idle}.

\vspace{0.02in}
\noindent\textbf{Android}
and Linux kernels are generally considered to have a significant ``gap'' because Android kernels are known to have its custom features designed for mobile devices (by heavily modifying the Linux kernel).
Intuitively, this should result in larger patch delays as Android kernel maintainers may need to maintain diverging copies of the same functionalities. However, we see that approximately 78\% of group-ported patches in Android are delayed for only one day.
The result may be counter-intuitive.
However, we note that a key maintainer of Android also serves as the main maintainer of the Linux LTS branches~\cite{android_common_repo,linux_stable_tree}, which may help explain the timeliness of the majority of patches.
%the gap between Linux and Android has been narrowing~\cite{anroid_back_to_mainline}.
We do note that the rest 22\% of patches incur significantly larger delays compared to most other distributions, likely due to the gap.
Finally, perhaps due to the same maintainer being involved in porting patches to both LTS and Android common, we observe fewer picked patches. In other words, the lack of maintainer diversity might have hurt Android common because there may be important mainline patches missed by LTS and also Android common.

\vspace{0.02in}
\noindent\textbf{Amazon}
uses minor version rebase to obtain patches from LTS.
As mentioned before, group-porting and minor version rebases achieve the same goal of getting patches from LTS.
This minor version rebase strategy is more feasible because Amazon has few picked patches from mainline. This makes it much easier because it just needs to port a small number of picked patches that are accumulated over time to the new base every time. 
The advantage of this strategy is that it enjoys a fairly small patch delays.
The weakness of this strategy is that Amazon will naturally have a lower patch rate, potentially missing some important mainline patches that should have been picked.

\cut{This is why they choose group-port instead where they would port LTS patches instead (if needed).}

\vspace{0.02in}
\noindent\textbf{Debian}
\phantomsection
\label{debian}
chooses to port a much smaller number of patches, i.e., very low patch rate, as mentioned before. Upon a closer inspection, we find that only when patches belonging to enabled kernel modules according to kernel configuration files are ported. As a result, its patch rate is the lowest. 
However, we argue this is generally a risky patch porting practice.
This is because configuration options can change over time and it can lead to a spike of workload to port patches pertaining to newly enabled modules, leading to sub-optimal results.
In addition, if others were to customize Debian by enabling additional modules, they will effectively end up with out-dated (and potentially more buggy) versions of those modules.

\begin{table}[]
\centering
\begin{tabular}{|r|rr|rr|}
\hline
\multicolumn{1}{|l|}{}          & \multicolumn{2}{c|}{In order commits}                                                   & \multicolumn{2}{c|}{Out-of order commits}                                               \\ \hline
\multicolumn{1}{|l|}{Linux LTS} & \multicolumn{1}{l|}{\begin{tabular}[c]{@{}l@{}}Fixes or \\ Cc tag\end{tabular}} & \multicolumn{1}{l|}{Big size} & \multicolumn{1}{l|}{\begin{tabular}[c]{@{}l@{}}Fixes or \\ Cc tag\end{tabular}} & \multicolumn{1}{l|}{Big size} \\ \hline
4.4 LTS                         & \multicolumn{1}{r|}{76\%}                                                    & 6.36\%                        & \multicolumn{1}{r|}{39\%}                                                    & 11.18\%                       \\ \hline
4.9 LTS                         & \multicolumn{1}{r|}{75\%}                                                    & 6.28\%                        & \multicolumn{1}{r|}{34\%}                                                    & 11.13\%                       \\ \hline
4.14 LTS                        & \multicolumn{1}{r|}{73\%}                                                    & 6.75\%                        & \multicolumn{1}{r|}{37\%}                                                    & 11.43\%                       \\ \hline
4.19 LTS                        & \multicolumn{1}{r|}{73\%}                                                    & 6.29\%                        & \multicolumn{1}{r|}{40\%}                                                    & 13.06\%                       \\ \hline
5.4 LTS                         & \multicolumn{1}{r|}{76\%}                                                    & 6.74\%                        & \multicolumn{1}{r|}{36\%}                                                    & 17.68\%                       \\ \hline
5.10 LTS                        & \multicolumn{1}{r|}{78\%}                                                    & 7.55\%                        & \multicolumn{1}{r|}{28\%}                                                    & 24.72\%                       \\ \hline
5.15 LTS                        & \multicolumn{1}{r|}{81\%}                                                    & 8.54\%                        & \multicolumn{1}{r|}{26\%}                                                    & 29.05\%                       \\ \hline
\end{tabular}
\caption{The comparison of in-order and out-of-order commits}
\label{table:inout_order}
\end{table}

\subsection{Factors Influencing the Results}

After the comparative study, we now dive deeper into the results and attempt to identify factors that influence the results.
We summarize them into a few points.

\vspace{0.1in}
\noindent\textbf{Lack of hints and large patches contribute to  the long patch delays from mainline to LTS.}
\begin{figure}
  \includegraphics[scale=0.44]{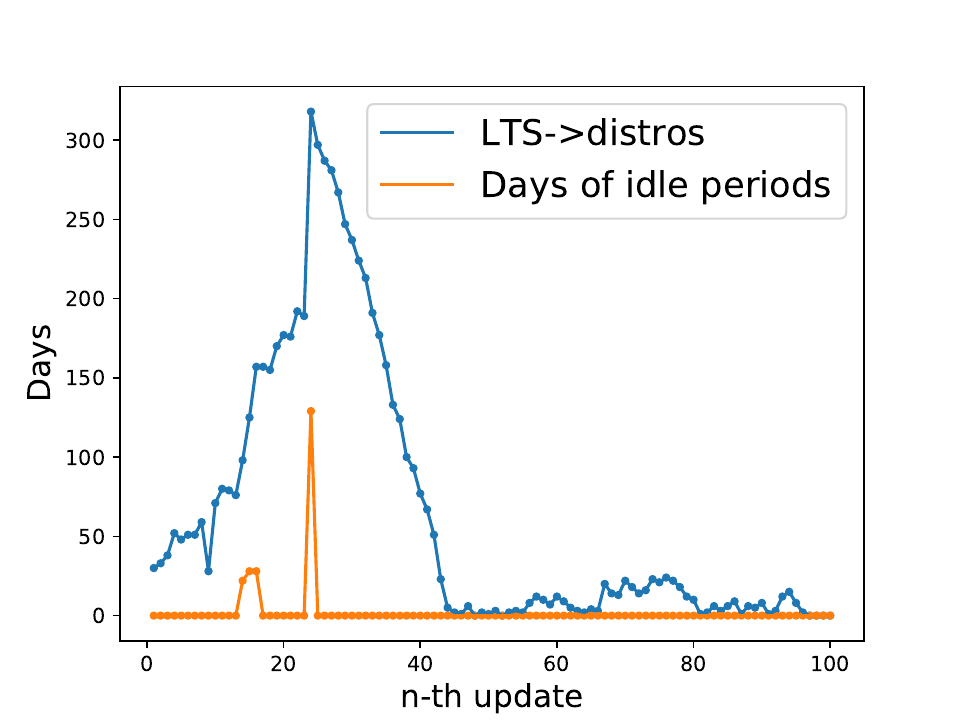}
  \caption{The delay changes for Ubuntu-18.04 over the time}
  \label{fig:portDelay_bionic}
\end{figure}
We know that the patch porting from Linux mainline to Linux LTS is an important step, as most distributions would have to wait for the patches to appear in LTS before they can port them. 
In general, we know it is a challenging task for LTS maintainers to identify and pick patches from mainline, as there are 
so many mainline commits that maintainers need to scan (a few hundred daily).
Therefore, it is possible that a mainline commit is mistakenly skipped and then picked up at a later point after the community realizes its importance.
We call such commits, out-of-order commits. To be more precise, commit A is "out-of-order" if there exists a later commit B in Linux mainline, whereas A is ported later to Linux LTS than B. If A is ported to Linux LTS before all later Linux mainline patches, B, we consider A to be ``in-order''.

According to our measurement of all LTS branches, we observe that surprisingly 30\% of patches are ported out-of-order. \autoref{fig:in_out_order_4_19} compares the delays for patches ported in-order vs. out-of-order. We can see clear
differences between in-order and out-of-order commits. All in-order commits are ported relatively quickly (under 80 days), whereas 50\% of out-of-order commits have delays longer than 100 days. In addition, about 78\% of out-of-order commits incur delays of 50+ days.

We explore the factors that could have led to the out-of-order commits.
As mentioned in \S\ref{tags}, there are two types of tags provided in the commits themselves that are designed to make the job of maintainers easier. Fixes tags can help maintainers make patch porting decisions. For instance, if a commit given in the fixes tag goes back to somewhere before Linux v4.4 LTS is forked, then we can infer that the v4.4 LTS branch would have inherited the bug,  therefore requiring the the patch. Cc stable tags indicate that downstream maintainers should pay more attention to the patch. Some CC stable tags also have additional kernel version indicators, such as \path{Cc: <stable@vger.kernel.org> # 4.14.x} and \path{Cc: stable@vger.kernel.org # v4.4+}, to suggest which LTS branch (\eg v4.14 and branches that after v4.4) may need the patch. The version tags can save time for maintainers.

According to various sources from maintainers~\cite{stable_kernel_process, sasha_lessons, sasha_stability} and our personal outreach to the Linux community, fixes tags and Cc stable tags are indeed consulted in practice when determining whether a mainline patch will be ported. 
We then measured the correlation between the presence of tags and the patching behaviors, as shown in ~\autoref{table:inout_order}. Indeed, there is a strong correlation. For those patches ported in-order, they are much more likely to contain either fixes or Cc stable tags; on the other hand, for out-of-order commits, the fractions with such tags are much lower. 

But there are still about 40\% of LTS commits that do not have any tags. According to maintainers~\cite{stable_kernel_process, sasha_lessons, sasha_stability}, commits without tags will go through the AUTOSEL bot, which is a neural network model introduced in Linux in 2017 to identify if a commit is a bug fix or feature commit; and then maintainers manually check whether the patch is necessary for specific Linux LTS branches. 
Since the AUTOSEL and the manual analysis by maintainers are blackboxes to us, we can only hypothesize what factors might influence the decision of whether to port a patch. In particular, we consider the patch size as a candidate factor. From the official document~\cite{stable_kernel_rules}, Linux LTS/stable branches do not accept patches larger than 100 lines. We therefore pick 75 lines as the threshold in determining ``big patches''. 
We hypothesize that larger patches are generally more complexity and more difficult to port, especially when the LTS/stable has diverged from the mainline. Indeed, \autoref{table:inout_order} shows that the percent of big patches in out-of-order commits are twice or three time as that in in-order commits(we also choose 50,60,70,80,90 lines as ``big patches'', but the result is the same).

\vspace{0.1in}
\noindent\textbf{Periods of inactivity of group-porting affects distribution's patch delays significantly.}
\phantomsection
\label{idle}
Now, we explore the patch porting behavior of distributions. 
As mentioned earlier, a primary source of patches is through group-porting from Linux LTS. 
It is natural to take advantage of the efforts by LTS maintainers who already determined which mainline patches are necessary to port for the base.
In other words, by following an LTS branch, downstream kernel can shield themselves from the large stream of mainline commits. 
\begin{table*}[]
\centering
\begin{tabular}{|l|r|r|rr|rrl|}
\hline
                          & \multicolumn{1}{l|}{}                                                                              & \multicolumn{1}{l|}{}                                                                                 & \multicolumn{2}{c|}{Picked \&  in upstream LTS}                          & \multicolumn{3}{c|}{Picked \& not in upstream LTS}                                                                                                                              \\ \cline{4-8} 
\multirow{-2}{*}{distributions} & \multicolumn{1}{l|}{\multirow{-2}{*}{\begin{tabular}[c]{@{}l@{}}\# of LTS\\ commits\end{tabular}}} & \multicolumn{1}{l|}{\multirow{-2}{*}{\begin{tabular}[c]{@{}l@{}}\# of picked\\ commits\end{tabular}}} & \multicolumn{1}{l|}{First in Distro} & \multicolumn{1}{l|}{First in LTS} & \multicolumn{1}{l|}{Total num} & \multicolumn{1}{l|}{\begin{tabular}[c]{@{}l@{}}\# of commits\\ with fix tags\end{tabular}} & \begin{tabular}[c]{@{}l@{}}\# of commits\\ necessary for LTS\end{tabular} \\ \hline
Oracle-Uek5 (v4.14 LTS)         & 25,663                                                                                              & 7,070                                                                                                  & \multicolumn{1}{r|}{570}             & 315                               & \multicolumn{1}{r|}{6,185}      & \multicolumn{1}{r|}{798}                                                                   & 192(24.1\%)                                                               \\ \hline
Oracle-Uek6 (v5.4 LTS)          & 22,069                                                                                              & 7,153                                                                                                  & \multicolumn{1}{r|}{220}             & 302                               & \multicolumn{1}{r|}{6,631}      & \multicolumn{1}{r|}{934}                                                                   & 200(21.4\%)                                                               \\ \hline
Oracle-Uek7 (v5.15 LTS)         & 13,790                                                                                               & 581                                                                                                   & \multicolumn{1}{r|}{78}              & 98                                & \multicolumn{1}{r|}{405}       & \multicolumn{1}{r|}{42}                                                                    & 14(33.3\%)                                                                   \\ \hline
Ubuntu-16.04 (v4.4 LTS)         & 18,974                                                                                              & 2,830                                                                                                  & \multicolumn{1}{r|}{226}             & 525                               & \multicolumn{1}{r|}{2,079}      & \multicolumn{1}{r|}{147}                                                                   & 39(26.5\%)                                                                \\ \hline
Ubuntu-18.04 (v4.14\&4.19 LTS)  & 50,771                                                                                             & 2,939                                                                                                  & \multicolumn{1}{r|}{569}             & 516                               & \multicolumn{1}{r|}{1,854}      & \multicolumn{1}{r|}{297}                                                                   & 104(35.0\%)                                                               \\ \hline
Ubuntu-20.04 (v5.4 LTS)         & 22,069                                                                                              & 12,67                                                                                                  & \multicolumn{1}{r|}{94}              & 149                               & \multicolumn{1}{r|}{1,024}      & \multicolumn{1}{r|}{122}                                                                   & 39(32.0\%)                                                                \\ \hline
Ubuntu-22.04 (v5.15 LTS)        & 13,790                                                                       & 859                                                                                                   & \multicolumn{1}{r|}{84}              & 183                                & \multicolumn{1}{r|}{592}       & \multicolumn{1}{r|}{51}                                                                    & 11(21.6\%)                                                                 \\ \hline
\end{tabular}
\caption{Comparison between picked commits and commits in corresponding LTS branches}
\label{table:comparison_pick_lts}
\end{table*}

As mentioned earlier and shown in ~\autoref{fig:groupport_delay},
Ubuntu and Oracle-Uek perform group-porting and they both have fairly poor delay profiles.
Interestingly, after we investigate the delay patterns, we observe some fairly anomalous data points.
Taking Ubuntu 18.04 as an example, 20\% of group-ported commits have a delay longer than 150 days. 
Furthermore, the long-delay commits are not uniformly distributed.
Instead, there is a strong temporal pattern as shown in ~\autoref{fig:portDelay_bionic}.
The X-axis shows the n-th group-porting events (each event corresponds to a group of commits) 
since the creation of the Ubuntu-18.04 branch.
The Y-axis shows the patch porting delay (for the line of ``LTS to distributions'')
and the number of days between two group-porting events (a.k.a., idle periods).
Upon a closer look, there are two clusters of ``idle periods''. 
The first cluster contains three small idle periods (around 25 days each).
The second contains a huge 144-day idle period where no group porting activity occurred.
These contribute to the massive delays in porting patches to Ubuntu, especially for the patches that have delays between 150 and 300 days.

To understand what happened behind the scene, we reached out to Ubuntu maintainers directly. Interestingly enough, 
we were told that those ``idle periods'' actually corresponded to a period of time where the company was busy with other important (but unspecified) tasks. 
According to our observation, during such periods, there seems to an increased focus on picking patches for CVEs (one possible task) — 1.7 CVE patches per day were picked during the idle period vs. 0.58 per day during the non-idle periods.

Surprisingly, we find that the idle periods are much longer in Oracle-Uek than Ubuntu: 
the longest idle periods in Oracle Uek-5, 6, and 7 are 233, 126, 112 days, respectively.
In contrast, the longest idle periods in Ubuntu-16.04, Ubuntu-18.04,  20.04, and 22.04 are 116, 144, 53, and 58 days, respectively.
This explains why the group-port delays (and the end-to-end patch delays) for Oracle-Uek are even worse than Ubuntu.

Note that SLE does not have the same problem because it group-ports patches only from a short-lived stable branch (for two to three months), so it is much less likely to incur such long idle periods.

\vspace{0.1in}
\noindent\textbf{Many patches in distributions should have been ported to LTS.}
\phantomsection
\label{distro_lts}
Since we know that distributions often pick patches directly from mainline and can have higher patch rate than LTS overall, does it mean that distributions actually offer better-quality kernels than LTS? Intuitively, that should be the case because it is the distribution kernels that are user-facing (not the LTS ones).
However, it would imply that there are patches missed or delayed by LTS (and they are necessary and should have been ported).

Indeed, as mentioned in \S\ref{sec:background}, distributions have their own bug tracking systems where they have external signals (e.g., from users) that trigger the patch picking. 
In addition, we find that CVE is also a focus for distributions (as they are responsible for the security of their customers).
In particular, maintainers usually attach a CVE ID to indicate that the patch fixes a known security vulnerability. 
Interestingly, we note that the picked CVE patches appear in distributions 74.2 days earlier than LTS on average; even if the picked CVE patches are later than LTS, it is only 16.7 days later on average.

To get a more comprehensive picture, \autoref{table:comparison_pick_lts} shows the picked commits by distributions and their relationship with LTS commits. We summarize three main points here.
First, the number of picked commits in distributions is generally much smaller than the number of Linux LTS commits. 
Second, we note that, in the cases where the picked commits  appear in both the LTS and distribution,  
many of the picked commits appear first in distributions and then in LTS, which is a strong evidence that distributions do pick useful patches earlier than LTS.
Finally, we note that a large fraction of picked commits are not even ported into LTS. 
We note that only a small fraction of them have tags, which is another evidence that that such pickings are driven by other external signals. 
More importantly, for patches with fixes tags, we find that roughly 20\% to 30\% of the patches should have been ported to LTS,
because the buggy commits (given in fixes tags) do appear in LTS, which shows that corresponding bugs are introduced in LTS, and LTS is required patches to fix those bugs.
Interestingly, the remaining 70\% to 80\% are cases where the buggy commits themselves are not in LTS. 
Upon a closer inspection, this is because the buggy commits were inherited by distributions exclusively through their picked patches.
This is the same phenomenon we reported earlier about higher bug inheritance ratio as a result of higher patch rate.

Overall, we conclude that many patches picked by distributions should have been ported to LTS as well.  In addition, if the patch delays
between mainline and LTS improve, that will benefit the entire ecosystem.

\section{Improvement of the Process}
\label{sec:suggest}

In this section, we summarize a few suggestions based on the insights gained from our measurement study.
To improve the patch porting process, we believe the responsibility is not only on Linux LTS maintainers but also the whole Linux community. Specifically, we propose the following suggestions.

\vspace{0.1in}
\noindent\textbf{Patch authors: increase the awareness and willingness to attach tags. }
According to the above analysis, tags associated with patch commits play a critical role in the patch porting process. 
The Cc stable tags and fixes tags directly inform LTS maintainers on whether a commit is a patch and a patch is necessary for stable or LTS branches. 
As we have shown, the lack of tags correlates directly with longer patch porting delays.
Interestingly, the patch submission in Linux is heavily decentralized. Anybody can submit patches 
and there is an official document outlining the desired patch submission process~\cite{patch_submit}, including the suggestion of including various tags. However, according to \cite{new_authors_num}, about 300 authors submit their first change into Linux mainline in every kernel release. They don't all know the rules.
This requires efforts to educate the developers and maintainers. For example, we note that sometimes patch reviewers will remind patch authors to add these tags but they are quite inconsistent.

Of course, generating these tags will requires some more effort on the patch authors, e.g., figuring out which commit introduced the bug. Therefore, it calls for automated reasoning tools that can generate these tags.

\vspace{0.1in}
\noindent\textbf{Maintainers: automated patch relevance analysis tools.}
For each Linux mainline commit, there are two indispensable steps for Linux LTS maintainers: (1) identify whether it is significant enough, e.g., an important bug fix, and conform to rules in \cite{stable_kernel_rules}; (2) check whether the commit should be ported into one or more specific stable/LTS branches.

Currently, there are already several machine learning tools~\cite{sasha_ML,hoang2019patchnet,hoang2020cc2vec} for the first step, the best accuracy is as high as 90\%. LTS maintainers also have already been using such tool, such as AUTOSEL~\cite{sasha_ML}.
\cut{
According to Sasha Levin's (one Linux LTS maintainer) presentations~\cite{sasha_ML}\cite{hoang2019patchnet}, he designed the feed-forward fully connected neural network with a set of manually selected features, such as the patch authors and the keywords in commit title, to identify the bug fix commits. Besides, there are other neural network research on this topic, such as Patchnet\cite{hoang2019patchnet} and CC2Vec\cite{hoang2020cc2vec}. The accuracies of these three are 81\%, 86\% and 90\% respectively. However, these tools are not widely used by Linux LTS maintainers. We compared the patch delay with Linux mainline between commits propagated by Sasha Levin and other Linux LTS maintainers. The result is shown in \autoref{fig:author_diff}, which shows that his tool is effective. We emailed with one Linux LTS maintainer--Ben hutchings, who told us that he did not use the Sasha Levin's machine learning tool.We also contacted the authors of Patchnet\cite{hoang2019patchnet} and CC2Vec\cite{hoang2020cc2vec}, they told us that they are used by maintainers. So we suggest the automated tools should be widespread for Linux LTS maintainers.}

Nevertheless, there are no automatic tools for the second step. The second step is required as not all stable/LTS branches may be affected by a bug. It is also more difficult. Linux LTS maintainer usually have to manually understand the patch and see if the patch is applicable to a specific stable/LTS branch. We believe the automation of this step can continue to help maintainers further.

\vspace{0.1in}
\noindent\textbf{Distribution maintainers: mitigating regression.}
As we see, a good fraction of patches ported by distributions (whether through LTS or picked directly from mainline) 
end up being buggy patches introducing new bugs. In other words, there are significant regression issues in patches for Linux mainline.
In particular, the bug inheritance ratio shows that a decent fraction came from picked patches.
Interestingly, we note that there is a chance to mitigate such issues.
The observation is that it is possible that the buggy patches already have fixes in Linux mainline when distributions picked the buggy patches themselves. Given that picking in generally is already substantially delayed, the likelihood of a patch for the initial picked patches being fixed on mainline is therefore quite high.
As a concrete example, in Oracle-Uek, we find that about 60\% - 70\% of buggy patches introduced by picking are fixed within one day.
This is likely because Oracle-Uek looked ahead to see if these picked patches on mainline have corresponding fixes already. If so, they ported those fixes as well.
However, other distros do not seem to have adopted this method. Using SLE as an example,
only 3.2\% - 6.4\% of picked buggy patches are fixed within a day.
We suggest this simple trick of look ahead on mainline to help mitigate the risk of regression caused by extensive picking (i.e., reducing the bug lifetime).

\vspace{0.1in}
\noindent\textbf{Community: cross validation metadata tracking system for all LTS and distributions.}
As shown in \autoref{table:comparison_pick_lts}, there are about 3.1\% - 19.4\% of picked commits that are ported into distributions before LTS.
And about 30\% of LTS-missed picked patches could possibly also be necessary for Linux LTS branches.
Also, the picking behaviors are drastically different even for same-based distributions.
Therefore, it is useful for distribution and LTS maintainers to review each other's patches to minimize missing patches. Furthermore, we suggest that the Linux community to collectively build a unified bug tracking system so that they can all benefit and quickly port important patches.

\section{Related Work}

\textbf{Linux patches measurement.} 
Zheng \etal~\cite{zhang2021investigation} have recently conducted a measurement on the Android kernel patch ecosystem, which is the most relevant work. They focus on the patch porting relationships in the Android branches, including measuring the patch porting delay, with an emphasis on analyzing binary-only OEM Android kernel images (e.g., Samsung). In our study, we focus on a very different target: downstream Linux distributions with public git repositories, without which it is difficult to infer the exact patch porting strategies. Finally, the measurement is limited to only CVE patches publicized on Android security bulletin, and focus on a single dimension of patch delay, whereas we additional derive insights from other metrics such as patch rate and bug inheritance ratio. 
Furthermore, their analysis for the patch delay lacks the knowledge of key data such as fixes tags and Cc tags, which we showed to be critical in assisting patch porting.

There are also other related research measuring patches and vulnerability life cycles. 
For example, Li \etal~\cite{li2017large} performed a large-scale measurement on CVE dataset across different various open-source projects, including Linux kernel. However, the study does not look at the Linux kernel as an ecosystem. Rather, the measurement focuses on Linux mainline only, e.g., when a bug was introduced in mainline and when it is fixed in mainline. 
Ozment \etal~\cite{ozment2006milk} measured whether security improves when software ages (using OpenBSD as a case study). 
Farhang \etal~\cite{farhang2019hey} measured the life cycle and timeline of patches in Android. Shahzadet \etal~\cite{shahzad2012large}  studied various dimensions of vulnerabilities such as risk level.
Finally, Jones \etal~\cite{jones2020deploying} conducted an extensive study on the impact manufacturers, carriers, and end-users have on the rollout of Android security updates and OS upgrades.

\vspace{0.1in}
\PP{Automated patch analysis.}
There exist several studies~\cite{hoang2019patchnet,wen2019ptracer,tian2012identifying} 
that use machine learning to determine whether a Linux mainline commit is a patch and should be ported to stable/LTS branches,
with varying degrees of success.

Alternatively, one can recognize whether a commit is a security patch and therefore should be considered for back-porting. For example, there have been studies using machine learning consuming commit messages and bug reports to achieve the goal~\cite{zhou2017automated,wijayasekara2014vulnerability,tyo2016empirical,goseva2018identification,behl2014bug,chaturvedi2012determining,menzies2008automated,ramay2019deep,roy2014towards}.
Another recent work by Wu \etal~\cite{wu2020precisely} takes a different route to analyze the bug semantics in a patch.

In addition, bug localization techniques\cite{rahman2018improving,zhou2012should,ye2014learning,wen2016locus,wang2014version,saha2013improving,lam2017bug,kim2013should,huo2016learning} can can automatically locate a potential bug in a software, which can help port more necessary patches. But they require bug reports which are unfortunately not available for the majority of the Linux mainline patches. Wen \etal~\cite{wen2019exploring} explored the relationships between bug introducing commits and bug fixing commits.

Finally, there are also recent work on determining whether a given patch has been applied to a specific target, for both source~\cite{mvp} and binary targets~\cite{fiber,jiang2020pdiff}. However, they cannot determine whether a patch really needs to be ported to the target.

\vspace{-.1in}

\vspace{0.1in}
\section{Conclusion}

In this work, we have performed a deep investigation into Linux kernel ecosystem to understand
the patch porting strategies in three metrics. We uncover many interesting findings on the current practices of patch porting,
including the causes of sub-optimal results. Finally, we also provide suggestions on how to improve the patch porting process.

\bibliography{reference}
\bibliographystyle{abbrv}

\end{document}